\documentclass[aps,prb,preprint,groupedaddress,superscriptaddress, graphicx, graphics, showpacs, ,showkeys]{revtex4-1}
\usepackage{amssymb}
\usepackage{amsmath}
\usepackage{amsfonts}
\usepackage{amstext}
\usepackage{amsgen}
\usepackage{amsbsy}
\usepackage{amsopn}
\usepackage{keyval}
\usepackage{trig}
\usepackage{graphicx}
\usepackage{graphics}
\usepackage{bm}
\usepackage{natbib}
\usepackage{textcase}

\begin{document}

\title{Revisiting the derivation of spin precession effects in quasi one-dimensional quantum wire models
}

\author{E. Papp}
\email[E. Papp]{, Email:  gerhardt_1916@yahoo.com}
\affiliation{Physics Department, West University of Timisoara, RO-300223, Timisoara, Romania}
\author{C. Micu}
\affiliation{Faculty of Science, North University of Baia Mare, RO-430122, Baia Mare, Romania}
\date{\today}
\begin{abstract}
In this paper one deals with the theoretical derivation of spin precession effects in quasi 1D quantum
wire models. Such models get characterized by equal coupling
strength superpositions of Rashba and Dresselhaus spin-orbit interactions of dimensionless magnitude $a$ under the
influence of in-plane
magnetic fields of magnitude $B$. Besides the wavenumber $k$ relying on the 1D electron, one accounts for
 the $s=\pm $  $1$ -
factors in the front of the square root term of the energy. Electronic structure properties of quasi 1D
semiconductor heterostructures like InAs quantum wires can then be readily discussed. Indeed, resorting to the
2D rotation matrix  provided by
competing displacements working along the Ox-axis opens the way to derive precession angles one looks for,
as shown recently. Proceeding further, we have to resort reasonably  to some extra conditions concerning the
general selection of the $k$-wavenumber via $kL=1$, where $L$ stands for the nanometer length scale of the quantum wire.
We shall also account for rescaled wavenumbers, which opens the
way to extrapolations towards imaginary and complex realizations. The parameter dependence of the precession
angles is characterized, in general, by interplays between admissible and forbidden regions, but large monotony
intervals are also in order.
\end{abstract}


\maketitle

\section{INTRODUCTION}

Spin-orbit interactions which are present in quasi one-dimensional (1D)
semiconductor heterostructures like $InAs$ quantum wires provide a promising
way to controllable spin manipulations\cite{1}. Besides the Rashba
spin-orbit interaction $V_{R}=\alpha _{R}\left( \sigma _{x}p_{y}-\sigma
_{y}p_{x}\right) /\hbar $ \quad\cite{2}, which is induced by an electric field,
one deals with the linearized Dresselhaus spin-orbit interaction $%
V_{D}=\alpha _{D}\left( \sigma _{x}p_{x}-\sigma _{y}p_{y}\right) /\hbar $ \quad
\cite{3}. These interactions rely on the presence of a structural and
crystal inversion symmetry, respectively. Superpositions of such
interactions under the influence of transversal\cite{4} and in-plane
magnetic fields\cite{5,6} like $\overrightarrow{B}=B(\cos \theta
,\sin \theta )$, have also been discussed. The polar angle between the
magnetic field and the $Ox$-axis is denoted by $\theta $. The corresponding
couplings are denoted by $\alpha _{R}$ and $\alpha _{D}$, in which case the
equal coupling strength  regime\cite{7,8} proceeds via $\alpha
_{0}=\alpha _{R}=\pm \alpha _{D}$. A typical choice is given by $\alpha
_{R}=2\times 10^{-11}eVm$, or equivalently, $a=1$, where $a=(\alpha
_{0}/2)\times 10^{11}/eVm$ denotes the dimensionless spin-orbit coupling.
Equal coupling strength superpositions of such interactions in the presence
of in-plane magnetic fields have also been analyzed recently\cite{9}. In
this later case it has been shown that accounting for spin conservations
amounts to the selection of the polar angle along the bisectrices. Having
obtained the energy band opens the way to the theoretical derivation of
novel spin precession effects, as shown by (72) in Ref. 9. For this purpose
one proceeds along the lines presented before\cite{10}, but now by resorting
to a different energy band structure. This amounts to reverse the usual $k$%
-wavenumber dependence of the energy in order to establish two correlated
wavenumbers, say $k_{+}$ and $k_{-}$, which are responsible for the
description of propagation paths along the $Ox$-axis. Then the two
dimensional rotation matrix\cite{11} one looks for can be established in
terms of displacements of length $L$ acting along the two paths just
referred to above\cite{9,10}. However, several details referring
to a systematic study of the parameter dependent spin precession angles
are still desirable. The suitable selection of the $k$-parameter deserves a
little bit more attention, too. Besides the influence of the discrete
parameter $s=\pm 1$, which is responsible for the $\pm $-signs in the front
of the square root of the energy, we shall account this time for a further
parameter, say $K$, reflecting the rescaling of squared wavenumber via $%
k^{2}\rightarrow Kk^{2}$. Handling the parameter dependence of spin
precession angles established in this manner amounts to deal with interplays
between $k$, $B$, $a$, $s$ and $K$, which represents our main motivation in
this paper. So far numerical $k$-inputs are introduced via $kL=1$, where $L$
denotes specifically the nanometer length scale of the quantum wire model.
Forbidden regions can be readily established in terms of selected parameters
for which the precession angles are imaginary. This happens in
configurations for which $s=-1$. Complementary intervals should then be
responsible for admissible configurations. The $K$-parameter opens the way
to extrapolations of the wavenumber towards imaginary and complex values,
which looks promising for further generalizations.

The paper is organized as follows. Preliminaries and notations are discussed
in Sec. II. The present Hamiltonian is introduced by neglecting, for
convenience, the orbital effects of the magnetic field. Accounting for spin
conservations leads to the selection of $\ $two $\theta $-angles, namely of $%
\theta =\pi /4$ and $\theta =3\pi /4,$ in which case $\alpha _{R}=\alpha _{D}
$ and $\alpha _{R}=-\alpha _{D}$ , respectively. Then the equal strength
limit of the energy can be readily established. This leads in turn to the
displacement momenta $k=$ $k_{\pm }$ serving to the description of two paths
one looks for, as indicated in Sec. III. In Sec. IV one shows that such
momenta work safely whenever $s=1$, but suitable convergence conditions have
to be accounted for in so far as $s=-1$. Precession angles established
before to first $\varepsilon $-order via $\Theta =(k_{+}-k_{-})L$ are
discussed in Sec. V. Numerical studies concerning these angles are presented
in some detail in Sec. VI. Section VII deals with imaginary and complex
realizations of the $k$-parameter. The Conclusions are presented in Sec.
VIII. Mathematical details concerning the derivation of precession angles
are shortly reviewed in the Appendix.

\section{PRELIMINARIES\ AND\ NOTATIONS}

In order to perform the quantum theoretical description of electronic
behavior in two-dimensional (2D) semiconductor heterostructures, single
particle Hamiltonians\cite{5,6}

\begin{equation}
H_{tot}=\frac{p_{x}^{2}+p_{y}^{2}}{2m^{\ast }}+V_{R}+V_{D}+V_{Z}+V\left(
y\right)
\end{equation}%
including the spin-orbit interactions have been proposed. This time the
orbital effects of the magnetic field have been neglected. The momentum
operator reads $\overrightarrow{p}=-i\hbar \nabla $, whereas $m^{\ast }$
stands for the effective mass of the electron. One has e.g. $m^{\ast
}=0.04m_{0}$ for $InAs$ quantum wires\cite{12}, where $m_{0}$ denotes the
usual rest-mass of the electron. The wire geometry is characterized by a
transversal confining potential, say $V\left( y\right) =m^{\ast }\omega
_{0}^{2}y^{2}/2$. The Zeeman interaction $V_{Z}=\mu _{B}\left( \sigma
_{x}B_{x}+\sigma _{y}B_{y}\right) g/2$ has also been incorporated, where $%
\mu _{B}=e\hbar /2m_{0}c$ stands for the Bohr-magneton, while $g$ denotes
the effective gyromagnetic factor. For convenience, we shall assume that $g=2
$, as usual.

Using the total Hamiltonian displayed in (1) leads to commutation relations
like

\begin{equation}
\left[ H_{tot},\sigma _{x}\mp \sigma _{y}\right] =\frac{2i\sigma _{z}}{\hbar
}\left( p_{x}\pm p_{y}\right) \left( \alpha _{R}\pm \alpha _{D}\right)
-i\sigma _{z}g\mu _{B}B\left( \sin \theta \mp \cos \theta \right) \quad
\end{equation}%
which exhibit the zero value if

\begin{equation}
\alpha _{R}=\mp \alpha _{D}\text{\qquad and\qquad }\tan \theta =\pm 1\quad .
\end{equation}%
So, one gets faced with conserved spin observables like $\sigma _{x}+\sigma
_{y}$ and $\sigma _{x}-\sigma _{y}$ when the Rashba and Dresselhaus
couplings exhibit the same magnitude, i.e. $\alpha _{R}=-\alpha _{D}$ and $%
\alpha _{R}=\alpha _{D},$ respectively. This proceeds in conjunction with
selected in-plane orientations of the magnetic field $\overrightarrow{B}$
for which $\tan \theta =1$ and $\tan \theta =-1$, respectively, as displayed
above.

The energy eigenvalue problem can be solved by resorting once more again to
the zero determinant condition for a homogenous system of two coupled
equations, as shown many times before\cite{13,14,15}. Starting from
the wavefunction

\begin{equation}
\Psi (x,y)=\exp (ikx)\Phi _{n}(y)\binom{\psi _{1}}{\psi _{2}}\quad
\end{equation}%
where $\Phi _{n}(y)$ stands for the oscillator eigenfunction, yields the
1D-reduction of (1) as

\begin{equation}
H_{1}=\frac{p_{x}^{2}}{2m^{\ast }}-\left( \alpha _{R}\sigma _{y}-\alpha
_{D}\sigma _{x}\right) \frac{p_{x}}{\hbar }+\frac{g}{2}\mu _{B}B(\sigma
_{x}\cos \theta +\sigma _{y}\sin \theta )+E_{n}^{0}\quad
\end{equation}%
in which the oscillator eigenvalue reads $E_{n}^{0}=\hbar \omega _{0}(n+1/2)$%
, as usual (see e.g.Ref. [13]). Now we are ready to apply the zero
determinant condition referred to above, which produces the energy

\bigskip
\begin{equation}
E=E_{n}^{(\pm )}(\theta )=\frac{\hbar ^{2}k^{2}}{2m^{\ast }}+E_{n}^{0}\pm
\sqrt{\Omega }
\end{equation}%
in accord with (16) in\cite{16}, where

\begin{equation}
\Omega =k^{2}(\alpha _{R}^{2}+\alpha _{D}^{2})+kg\mu _{B}B\left( \alpha
_{D}\cos \theta -\alpha _{R}\sin \theta \right) +\left( \frac{g}{2}\mu
_{B}B\right) ^{2}\quad .
\end{equation}%
and where the $\pm $-signs in the front of $\sqrt{\Omega }$ can also
be viewed, in general, as reflecting the influence of the spin\cite{16,17}. We then have to introduce a further parameter like $s=\pm 1$,
which will be used hereafter. Accordingly, the lower energy configuration
corresponds to $s=-1$.

The corresponding spinorial eigenfunction is given by $\psi _{1}=1/\sqrt{2}$
 and
\linebreak $\psi _{2}=\pm \exp (-i\beta )/\sqrt{2}$, where

\begin{equation}
\tan \beta =\frac{k\alpha _{R}-(g/2)\mu _{B}B\sin \theta }{k\alpha
_{D}+(g/2)\mu _{B}B\cos \theta }\quad .
\end{equation}%
The equal strength limit of (8) can also be readily performed. One would
then obtain $\tan \beta =-\tan \theta $, in accord with (3), so that $\beta
=\pi -\theta $. This means that $\beta =\pi /4$ $(3\pi /4)$ if $\alpha
_{R}=\alpha _{D}$ $(\alpha _{R}=-\alpha _{D})$.

Accounting for the spin conservation, we have to realize that the
symmetrized equal coupling strength limit of the present spin dependent but
non-symmetrized energy band (6) is given by

\begin{equation}
E_{n}^{(s)}(B,a)=\frac{\hbar ^{2}k^{2}}{2m^{\ast }}+E_{n}^{0}+\frac{s}{2}%
\sqrt{8k^{2}\alpha _{0}^{2}+\left( g\mu _{B}B\right) ^{2}}
\end{equation}%
by virtue of (3). So far, the dimensionless spin-orbit coupling, i.e. $%
a=(\alpha _{0}/2)\times 10^{11}/eVm$ , has the magnitude order of unity. One
sees that the linear $B$-dependent term under the square root in (6) is
ruled out, which comes definitely from the inter-related equal coupling
strength limit one deals with in this paper. Moreover, ruling out the $B$%
-dependent term leads to the conversion of (6) into a conditionally solvable
biquadratic equation in $k^{2}$, which serves as a starting point to the
derivation of spin precession effects.

Rescaled variables like $\widetilde{\alpha }_{0}=m^{\ast }\alpha _{0}/\hbar
^{2}$, $\widetilde{B}=m^{\ast }B/\hbar ^{2}$, $E_{n}^{(s)}(B,\alpha
_{0})=\hbar ^{2}\xi ^{2}/2m^{\ast }$ and $E_{n}^{0}=\hbar ^{2}\kappa
_{n}^{2}/2m^{\ast }$ can also be introduced. Numerical studies can then be
readily done by starting from $m^{\ast }/\hbar ^{2}\cong 5.24936\times
10^{17}/eVm^{2}$, $g\mu _{B}\cong 1.157668\times 10^{-4}eV/T$, $g\mu
_{B}m^{\ast }/\hbar ^{2}\cong 6.077016\times 10^{13}/m^{2}T$ , $\widetilde{%
\alpha }_{0}\cong 1.049872a\times 10^{7}/m$ and $n=0$. We shall also assume
that $E_{0}^{0}=1meV$ \quad \cite{13}. We have to keep in mind that present
calculations are sensitive to the numerical selection of $k$-parameter. We
have to realize that such selections can be established reasonably via$\ kL=1
$, which yields a typical nanometer length like $L=10^{-7}m$, when $%
k=10^{7}/m$. However, we have to be aware that other selections, $k$%
-dependent ones included, are conceivable. For convenience, we shall insert
hereafter the wavenumber input $k=10^{7}/m$, though the choice $k=10^{8}/m$
has been used tentatively before\cite{9}.
\section{INVERTING\ THE\ WAVENUMBER\ DEPENDENCE\ OF\ THE\ ENERGY}

It is also clear that (9) can be rewritten equivalently as

\begin{equation}
\xi ^{2}-\kappa _{n}^{2}=A_{s}(B,\alpha _{0},k)=A_{\pm }(B,\alpha
_{0},k)\equiv k^{2}+s\left[ 8k^{2}\widetilde{\alpha }_{0}^{2}+(g\mu _{B}%
\widetilde{B})^{2}\right] ^{1/2}
\end{equation}%
where $s=\pm 1$, which produces the factorized algebraic equation

\begin{equation}
k^{4}-2(A_{\pm }+4\widetilde{\alpha }_{0}^{2})k^{2}+A_{\pm }^{2}-(g\mu _{B}%
\widetilde{B})^{2}=(k^{2}-k_{+}^{2})(k^{2}-k_{-}^{2})
\end{equation}%
where $k^{2}=k_{\pm }^{2}$ stand implicitly for the roots serving to the
description of two propagation paths. Next we shall resort to a further
approximation, namely to handle in the sequel the $A_{\pm }$-functions in
terms of numerical inputs for the $k$-parameter. Then we are in a position
to establish actually $k$-roots one looks for as

\begin{equation}
k=k_{\pm }=\Omega _{0}\sqrt{1\pm \varepsilon }
\end{equation}%
where

\begin{equation}
\varepsilon =\frac{\left[ 8\widetilde{\alpha }_{0}^{2}A_{\pm }+16\widetilde{%
\alpha }_{0}^{4}+(g\mu _{B}\widetilde{B})^{2}\right] ^{1/2}}{\Omega _{0}^{2}}%
>0
\end{equation}%
and

\begin{equation}
\Omega _{0}=\left( 4\widetilde{\alpha }_{0}^{2}+A_{\pm }\right) ^{1/2}\quad .
\end{equation}%
The same job can be done by resorting to numerical energy-inputs, such as
indicated by the equation

\begin{equation}
E_{n}^{(1)}+E_{n}^{(-1)}-2E_{n}^{0}=\frac{\hbar ^{2}k^{2}}{m^{\ast }}\quad
.\qquad
\end{equation}%
Accordingly, one obtains $k=10^{7}/m$ when $%
E_{0}^{(1)}+E_{0}^{(-1)}=2.1905meV$.

\section{CONVERGENCE\ CONDITIONS}

One remarks that (12) produces a power series in terms of the convergence
condition $\varepsilon ^{2}<1$, which is synonymous to

\begin{equation}
g\mu _{B}\mid \widetilde{B}\mid <\mid A_{\pm }\mid \quad .
\end{equation}%
This inequality is fulfilled automatically if $s=1$. However one gets faced
with extra conditions like

\begin{equation}
8\widetilde{\alpha }_{0}^{2}-k^{2}>2g\mu _{B}\mid \widetilde{B}\mid >0
\end{equation}%
if $s=-1$, provided that

\begin{equation}
8\widetilde{\alpha }_{0}^{2}>k^{2}\quad .
\end{equation}%
Next, we shall handle (17) via $8\widetilde{\alpha }_{0}^{2}=0(1)$, $2g\mu
_{B}\mid \widetilde{B}\mid =0(1)$ and $k^{2}=0(\epsilon )$, where $\epsilon $
stands for a small parameter. Such conditions are reminiscent to the
asymptotic description of nonlinear oscillations\cite{18}. The interesting
point is that (17) provides an admissible but finite $B$-interval such as
given by

\begin{equation}
B\in I_{B}(a)=\left( -B_{c}a^{2},B_{c}a^{2}\right)
\end{equation}%
in so far as the $k^{2}$-term is neglected, where

\begin{equation}
B_{c}=\frac{16m^{\ast }}{g\mu _{B}\hbar ^{2}}10^{-22}T\quad .
\end{equation}%
It should be noted that $a$ plays this time the role of an input parameter.
Conversely, starting from an input $B$-parameter leads to two disjoint but
admissible semi-infinite $a$-intervals like

\begin{equation}
\widetilde{I}_{a}^{(-)}(B)=\left( -\infty ,-\sqrt{\frac{B}{B_{c}}}\right)
\end{equation}%
and

\begin{equation}
\widetilde{I}_{a}^{(+)}(B)=\left( \sqrt{\frac{B}{B_{c}}},\infty \right)
\end{equation}%
so that (18) gets converted into

\begin{equation}
a\in \widetilde{I}_{a}(B)=\widetilde{I}_{a}^{(-)}(B)\cup \widetilde{I}%
_{a}^{(+)}(B)\quad .
\end{equation}%
Concrete realizations concerning (19) and (23) will be presented below.

Under such conditions the leading approximations characterizing (12) are
given by

\begin{equation}
k=k_{\pm }=\Omega _{0}\left( 1\pm \frac{\varepsilon }{2}\right) =\Omega
_{0}\pm \Omega _{1}
\end{equation}%
both for $s=1$ and $s=-1$, where $\Omega _{1}=\varepsilon \Omega _{0}/2$,
with the understanding that in the latter case the wavenumber description
proceeds in terms of (19) and (23).

\section{SPIN\ PRECESSION\ EFFECTS}

Displacements of length $L$ along the $Ox$-axis can be readily applied by
resorting to the orthonormalized spinor

\begin{equation}
\left\vert \Psi _{\pm }\right\rangle =\frac{1}{\sqrt{2}}\left(
\begin{array}{c}
1 \\
\pm \exp \left( -i\beta \right)%
\end{array}%
\right)
\end{equation}%
by virtue of (4). This results in a 2D rotation matrix [10,11], providing in
turn the precession angle as

\begin{equation}
\Theta (B,a;s)=2\Omega _{1}L=\frac{\alpha _{0}m^{\ast }L}{\hbar ^{2}}\left[
\frac{16\alpha _{0}^{2}\left( m^{\ast }/\hbar ^{2}\right) ^{2}+\left( g\mu
_{B}B/\alpha _{0}\right) ^{2}+8A_{\pm }}{4\alpha _{0}^{2}\left( m^{\ast
}/\hbar ^{2}\right) ^{2}+A_{\pm }}\right] ^{1/2}
\end{equation}%
as shown in the Appendix, which proceeds in accord with (72) in Ref. 9. Our
main emphasis in this paper is one Refs. 9 and 10, but we have to mention
that starting with the idea of the voltage controlled spin precession\cite{19}, a multitude of spin precession descriptions have also been presented
during time\cite{20,21,22,23,24}, numerical studies included\cite{25,26}.

Now we are ready to discuss the parameter dependence of the precession angle
in a more systematic manner. Inserting $k=10^{7}/m$ , provides in turn the $%
B $- and $a$-dependence of the precession angle $\Theta (B,a;s)$ in terms of
$a $- and $B$-inputs respectively. One finds that $\Theta (B,a;s)$ is a
positive concave function \ of $B$ and $a$ increasing monotonically with $%
\mid B\mid $ and $\mid a\mid $, respectively, in so far as $s=1$. However,
one gets faced with nontrivial patterns when $s=-1$. So there are crossing
points between $\Theta $-plots for $s=1$ and $s=-1$ concerning both $B$- and
$a$-dependent curves. Such points are located at

\begin{equation}
B=B_{\pm }(a)=\pm B_{c}a^{2}
\end{equation}%
and

\begin{equation}
a=a_{\pm }(B)=\pm \sqrt{\frac{B}{B_{c}}}
\end{equation}%
respectively. These points reproduce identically the edge points
characterizing the admissible intervals (19) and (23). It should be noted
that both (27) and (28) are produced by the basic equation

\begin{equation}
(g\mu _{B}\widetilde{B})^{2}=16\widetilde{\alpha }_{0}^{4}
\end{equation}
which has the meaning of a leading approximation. The upper bounds
characterizing the precession angle within the admissible intervals are then
given by

\begin{equation}
\Theta =\Theta _{1}(a)=\Theta (B_{\pm }(a),a;s=-1)=2.099744\sqrt{2}a
\end{equation}%
and

\begin{equation}
\Theta =\Theta _{2}(B)=\Theta (B,a_{\pm }(B);s=-1)=2.099744\sqrt{\frac{2B}{%
B_{c}}}
\end{equation}%
where $a$ and $B$ stand for inputs. Other characteristic point of interest
are the zeros exhibited by the $s=-1$-precession angle on the $B$- and $a$%
-axes. One realizes that such zeros are inter-related with the onset of
discontinuity points of \ the second kind such that $\Re\Theta
(B,a;s=-1)=0$, as displayed in Figs. 1-4. Handling this latter equation gives

\begin{equation}
k^{2}+\sqrt{k^{4}+(g\mu _{B}\widetilde{B})^{2}}=4\widetilde{\alpha }_{0}^{2}
\end{equation}%
which yields the solutions

\begin{equation}
B=B_{\pm }^{(0)}(a)=\pm \frac{2\sqrt{2}\widetilde{\alpha }_{0}}{g\mu
_{B}m^{\ast }/\hbar ^{2}}\sqrt{2\widetilde{\alpha }_{0}^{2}-k^{2}}\in
I_{B}(a)
\end{equation}%
or

\begin{equation}
a=a_{\pm }^{(0)}(B)=\pm \frac{\hbar ^{2}}{4m^{\ast }}\left[ k^{2}+\left[
k^{4}+(g\mu _{B}\widetilde{B})^{2}\right] ^{1/2}\right] ^{1/2}\in \widetilde{%
I}_{a}(B)
\end{equation}%
respectively. It can be easily verified that the zeros established in this
manner get included into the admissible intervals (19) and (23). We have to
remark that the that the zeros of the denominator in (31) such as given by

\begin{equation}
(g\mu _{B}\widetilde{B})^{2}=16\widetilde{\alpha }_{0}^{4}+k^{4}
\end{equation}%
produce vertical singularity lines which are asymptotically
indistinguishable from supercritical $\Theta $-tails for which $\Theta
(B,a;s=-1)\eqslantgtr \Theta _{1}$ and $\Theta (B,a;s=-1)\eqslantgtr \Theta
_{2}$, which proceed in connection with $B=B_{\pm }(a)$ and $a=a_{\pm }(B)$,
respectively. The understanding is that (29) is a leading approximation
which comes from (35) via $k^{2}\rightarrow 0$. Such vanishingly small
strips tails lying outside admissible regions should then be viewed as
meaningless artifacts which can be hereafter ignored. Under such conditions
one gets faced with a leading approximation for which $\Theta
(B,a;s=1)\eqslantgtr \Re\Theta (B,a;s=-1)$, which opens the way to a
reasonable synthesis of admissible $\Theta $-trajectories in pertinent
parameter spaces.

After having been arrived at this stage, we are in a position to say that
forbidden regions should proceed complementarily in terms of selected
parameters for which the imaginary parts of precession angles are non-zero,
i.e. for $\Im\Theta (B,a;s=-1)\neq 0$. It is understood that this
latter inequality proceeds in combination with $\Re\Theta (B,a;s=-1)=0$%
. Such parameters belong to intervals like

\begin{equation}
C_{B}(a)=(-\infty ,B_{-}(a))\cup (B_{+}(a),\infty )
\end{equation}%
and

\begin{equation}
\widetilde{C}_{a}(B)=(a_{-}(B),a_{+}(B))
\end{equation}%
which are complementary to (19) and (23), respectively.

\section{NUMERICAL\ STUDIES}

Concrete plots displaying the parameter dependence of precession angles are
presented in Figs. 1-4. We have to anticipate that in all these cases the
precession angles are positive whenever $s=1$. The precession angle is
presented in Fig. 1 for $a=1$, $s=1$ (solid curve) and $s=-1$ (dashed
curve). The crossing points are located at $B_{\pm }(1)=\pm B_{c}\cong \pm
7.255082T$, while the zeros are given by $B_{\pm }^{(0)}(1)\cong \pm
5.363195T$. The admissible $B$-interval is given by $I_{B}(1)$, while the
forbidden one proceeds via $C_{B}(1)$, in accord with the dotted curve in
Fig. 1. The horizontal and vertical dotted lines serve to the discrimination
of crossing points, and the same concerns Figs. 2-4. Inserting $a=2$ instead
of $a=1$ leads to similar patterns, as shown in Fig. 2. The crossing points
and the zeros are now given by $B_{\pm }(2)\cong 29.020328T$ and $B_{\pm
}^{(0)}(2)\cong 27.326675T$, respectively. The admissible interval concerns
this time $I_{B}(2)$, which is complementary to the forbidden interval $%
C_{B}(2)$ in which $\Im\Theta (B,2;s=-1)\neq 0$.

\begin{figure}[ht]
\includegraphics[scale=0.67,clip=]{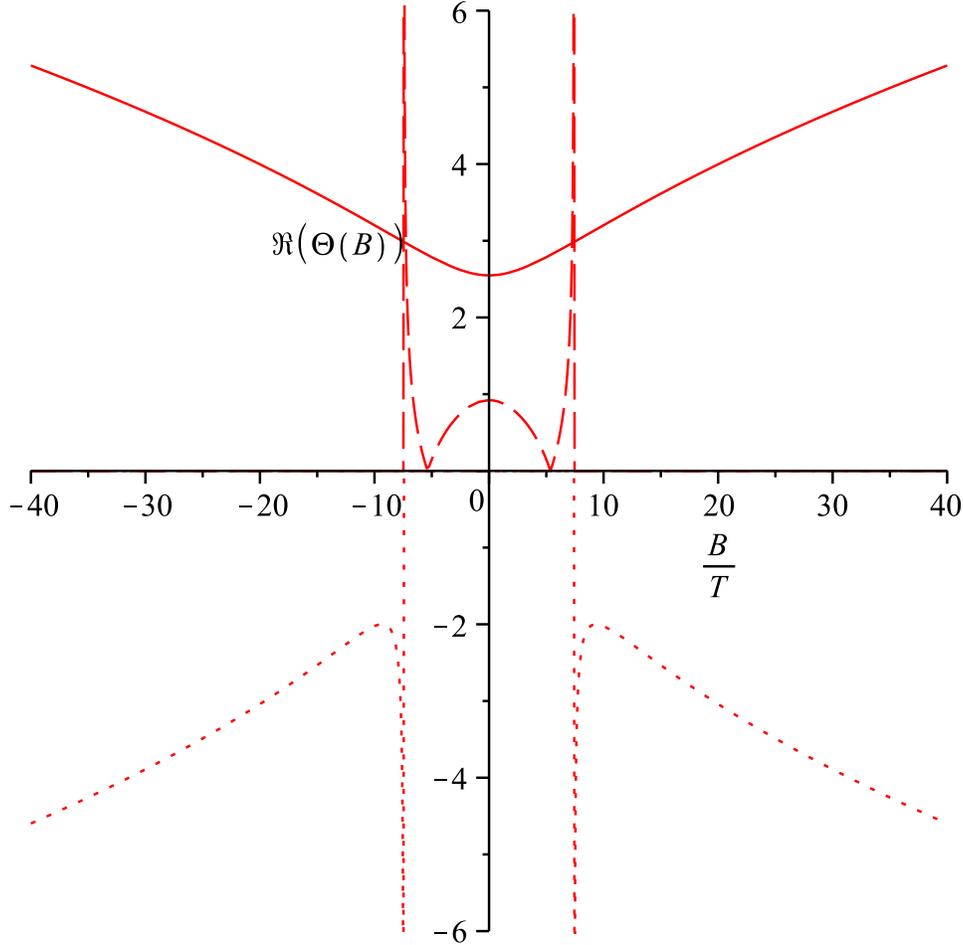}
\caption{\label{FIG_1_Papp}The $B$-dependence of the real parts of precession angles for $a=1$,
$s=1$ (solid curve) and $s=-1$(dashed curve). Dotted curves displaying the $%
B $-dependence of $\Im\Theta (B,a=1;s=-1)$ have also been inserted.}
\end{figure}

\begin{figure}[ht]
\includegraphics[scale=0.67,clip=]{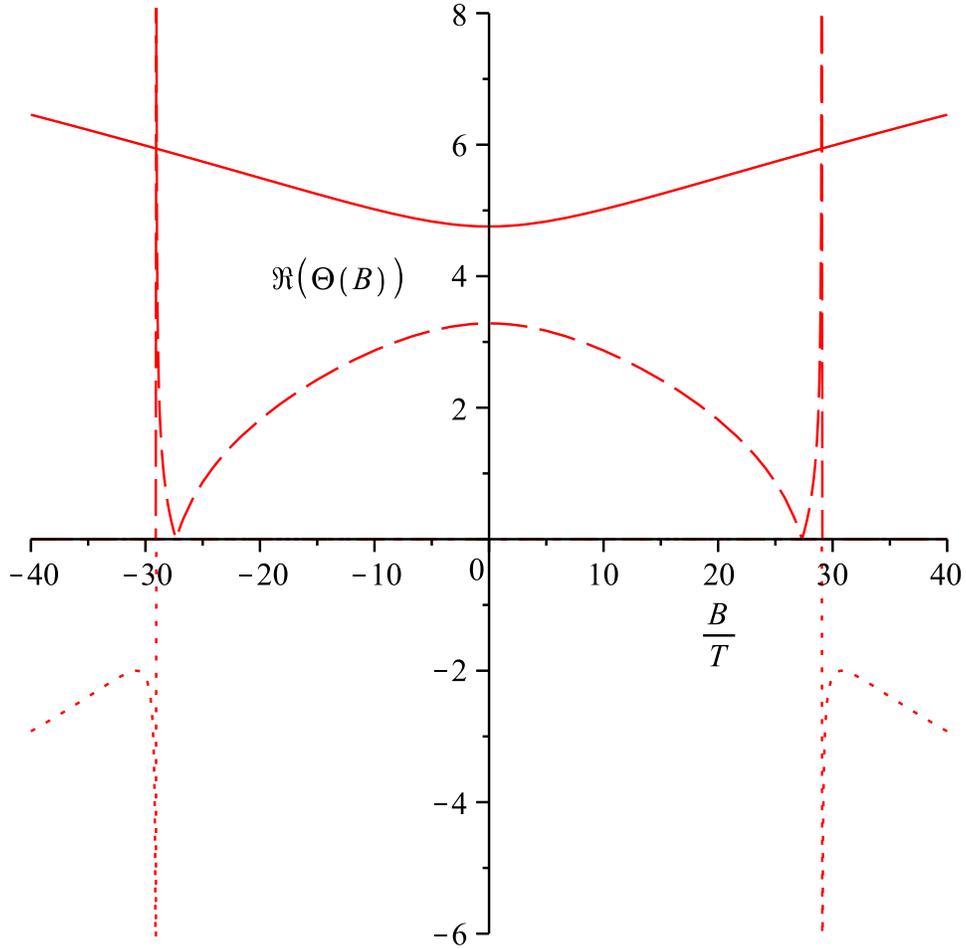}
\caption{\label{FIG_2_Papp}The $B$-dependence of the real parts of precession angles for $a=2$,
$s=1$ (solid curve) and $s=-1$(dashed curve). Now the dotted curves concern
the $B$-dependence of $\Im\Theta (B,a=2;s=-1)$.}
\end{figure}

The $a$-dependence of the precession angle can be discussed in a rather
similar manner. Inserting $B=6T$, leads to typical plots for $s=1$ (solid
curves) and $s=-1$ (dashed curves), which exhibit the crossing points $%
a_{\pm }(6)\cong \pm 0.827006$ and the zeros $a_{\pm }^{(0)}(6)\cong \pm
1.041294$, as shown in Fig. 3. However, this time the admissible interval is
expressed by two disjoint semi-infinite intervals, i.e. by $\widetilde{I}%
_{a}(6)$. The dotted curve concerns this time the forbidden interval $%
\widetilde{C}_{a}(6)$. The $B=12T$- counterparts of these plots are
displayed in Fig. 4. Just note that in this latter case crossing points and
zeros are given by $a_{\pm }(12)\cong \pm 1.286084$ and $a_{\pm
}^{(0)}(12)\cong \pm 1.377026$.

\begin{figure}[ht]
\includegraphics[scale=0.67,clip=]{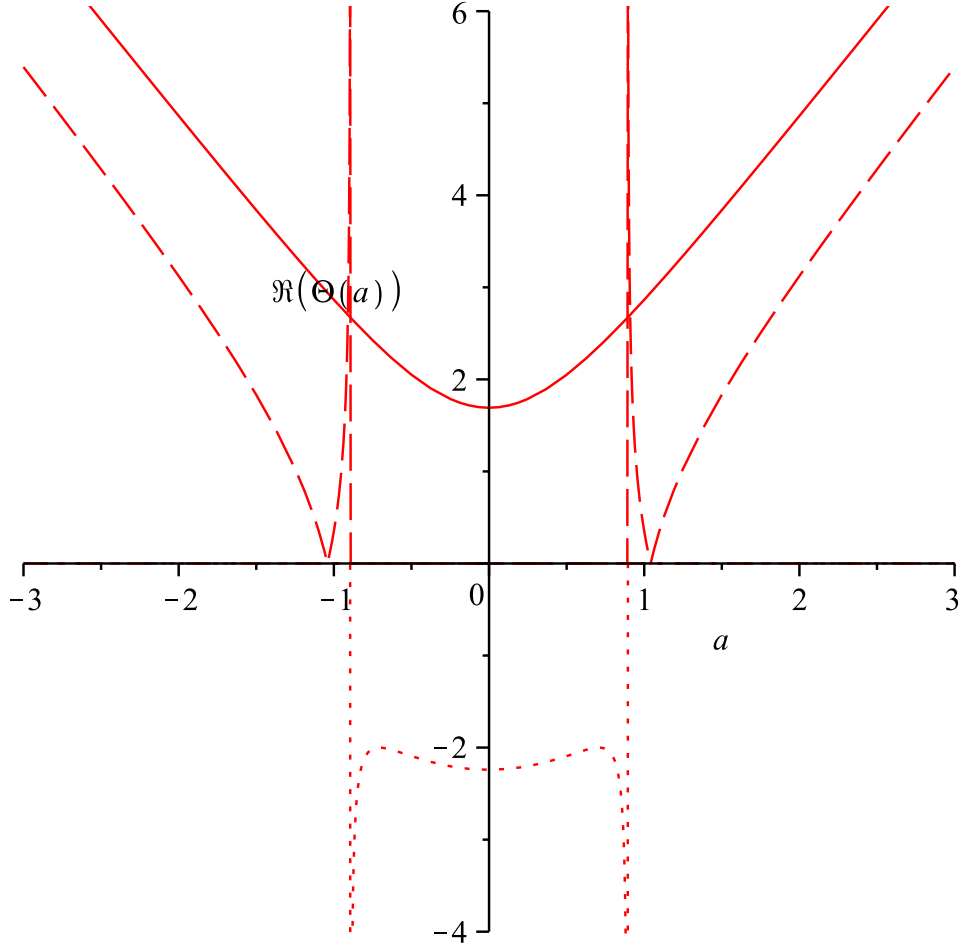}
\caption{\label{FIG_3_Papp}The $a$-dependence of the real parts of precession angles for $B=6T$%
, $s=1$ (solid curve) and $s=-1$ (dashed curve). The $a$-dependence of $%
\Im\Theta (B=6,a;s=-1)$ is displayed by the dotted curve.}
\end{figure}

\begin{figure}[ht]
\includegraphics[scale=0.67,clip=]{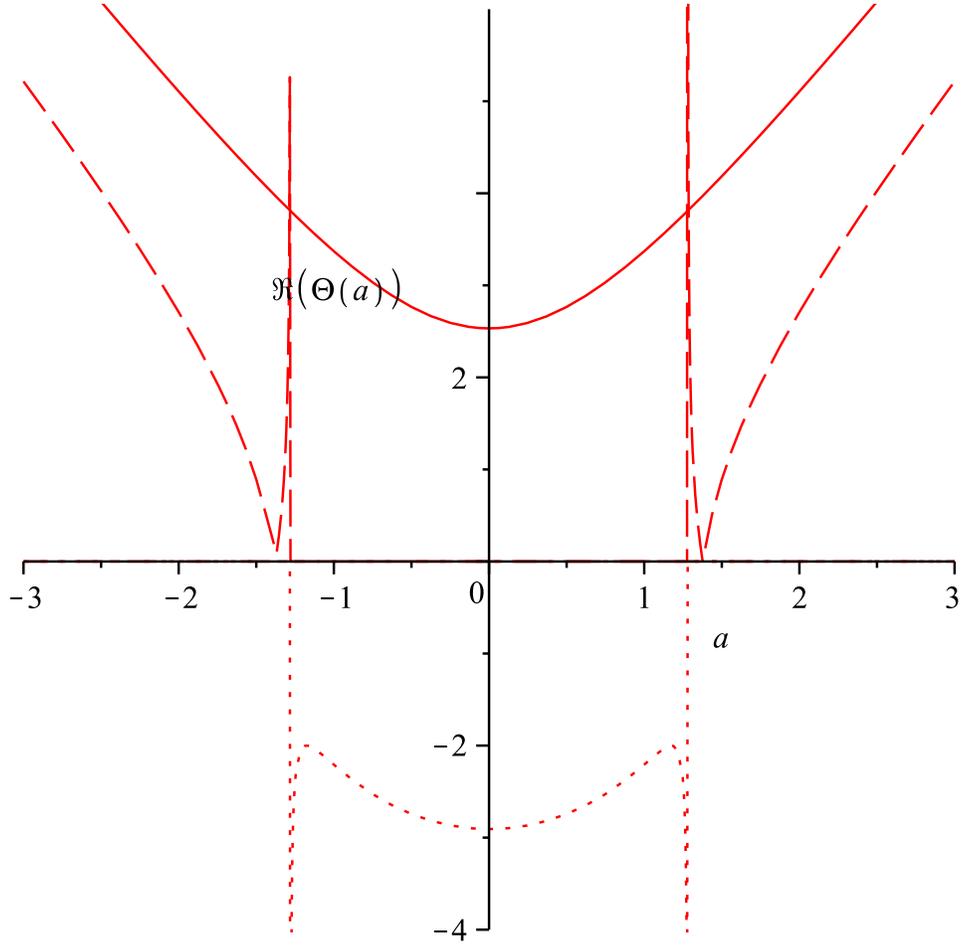}
\caption{\label{FIG_4_Papp}The $a$-dependence of the real parts of precession angles for $B=12T$%
, $s=1$ (solid curve) and $s=-1$ (dashed curve). Now the dotted curve is
responsible for the $a$-dependence of $\Im\Theta (B=12,a;s=-1)$.
}
\end{figure}


\section{THE\ INFLUENCE\ OF\ IMAGINARY\ AND\ COMPLEX\ WAVENUMBERS}

Next let us rescale the squared wavenumber as

\begin{equation}
k^{2}\rightarrow k_{R}^{2}=Kk^{2}
\end{equation}%
which means that imaginary $k-$realizations proceed via $K<0$. The $K$%
-dependence of $\Theta =\Theta (B,a;s,K)$ can be easily displayed, as shown
in Fig. 5 for $B=6T$ and $a=1$. One remarks that the imaginary parts of
precession angles, such as indicated by dot-dashed ($s=1$) and dotted ($s=-1$%
) curves, are zero, unless $K<K_{c}$, where $K_{c}\cong -1.51$.
Correspondingly, the real parts of present $\Theta $-angles, which are
displayed by solid ($s=1$) and dashed ($s=-1$) curves, get characterized by
a discontinuity point of second kind for which $\Re\Theta (6,1;s=\pm
1,K_{c})$ $\cong 2.549063$. The imaginary and real parts obey the symmetry
properties

\begin{equation}
\Im(\Theta (6,1;s=1,K)+\Theta (6,1;s=-1,K))=0
\end{equation}%
which is valid irrespective of $K$ and

\begin{equation}
\Re\Theta (6,1;s=1,K)=\Re\Theta (6,1;s=-1,K)
\end{equation}%
working for $K\eqslantless K_{c}$. A further discontinuity point of second
kind is located at $K=K_{1}\cong 0.692018$, such that $\Theta
(6,1;s=-1,K_{1})=0$, which is similar to the zeros displayed by $\Re
\Theta (B,a;s=-1,K=1)$ in Figs. 1-4.

\begin{figure}[ht]
\includegraphics[scale=0.67,clip=]{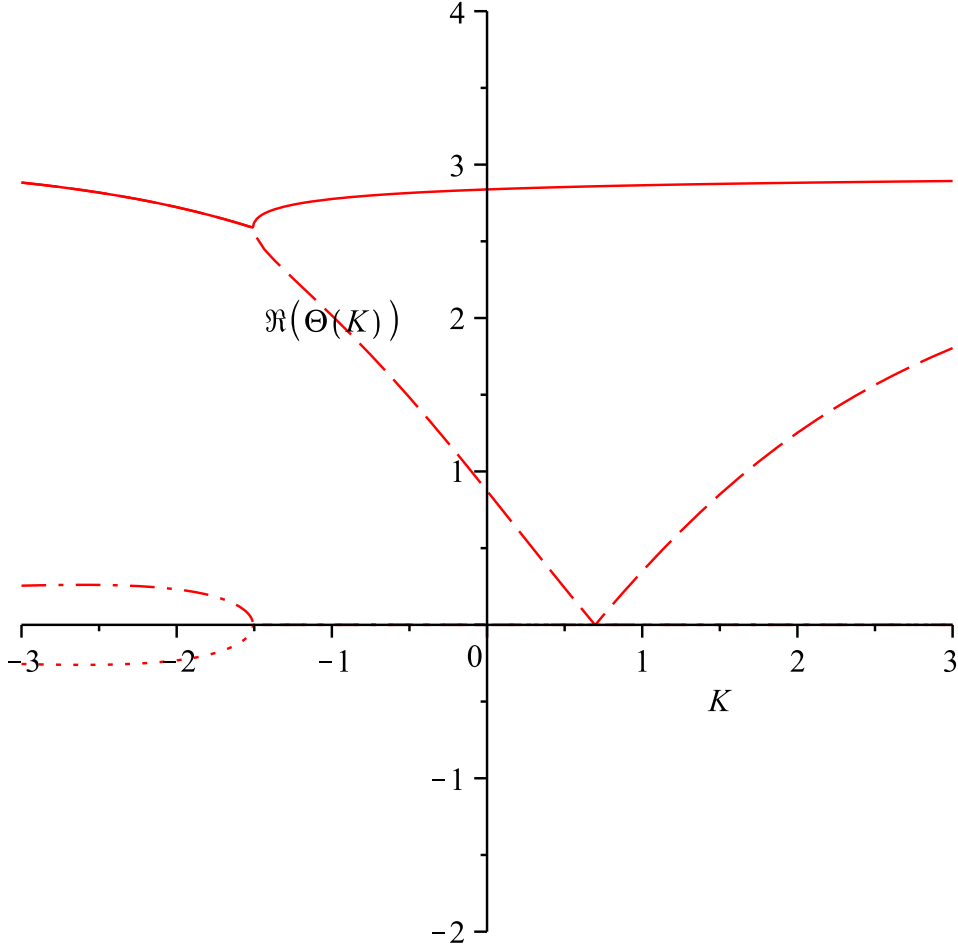}
\caption{\label{FIG_5_Papp}The $K$-dependence of the real parts of precession angles for $B=6T$%
, $a=1$, $s=1$ (solid curve) and $s=-1$ (dashed curve). The $K$-dependence
of $\Im\Theta (B=6,a=1;s=\pm 1,K)$ \ has also been included, as shown
by dot-dashed ($s=1$) and dotted ($s=-1$) curves.
}
\end{figure}

A further interesting example concerns the $a$-dependence of real parts of
precession angles for $B=6T$ and $K=i$, as shown by solid ($s=1$) and dashed
($s=-1$) curves in Fig. 6. Dot-dashed ($s=1$) and dotted ($s=-1$) curves,
which are responsible for the imaginary parts, have again been inserted. It
is clear that now the rescaled wavenumber exhibits the complex form $%
k_{R}=\exp (i\pi /4)k$. Just remark the symmetry of present numerical
realizations such as given by

\begin{equation}
\Theta (B=6,a=0;s=1,K=i)=1.858427-i0.250224
\end{equation}%
and

\begin{equation}
\Theta (B=6,a=0;s=-1,K=i)=-i1.858427+0.250224
\end{equation}%
Vertical asymptotes reflecting the influence of crossing points $a=a_{\pm
}(B=6)\cong 0.909399$ discussed before by virtue of (23) can also be readily
identified. The present patterns are similar to the ones displayed in Fig.
4, but now one deals with additional imaginary contributions concerning $%
\Im\Theta (B=6,a;s=1,K=i)$ as well as $\Im\Theta
(B=6,a;s=-1,K=i) $, this time for $a\in \widetilde{I}_{a}(B=6)$.

\begin{figure}[ht]
\includegraphics[scale=0.67,clip=]{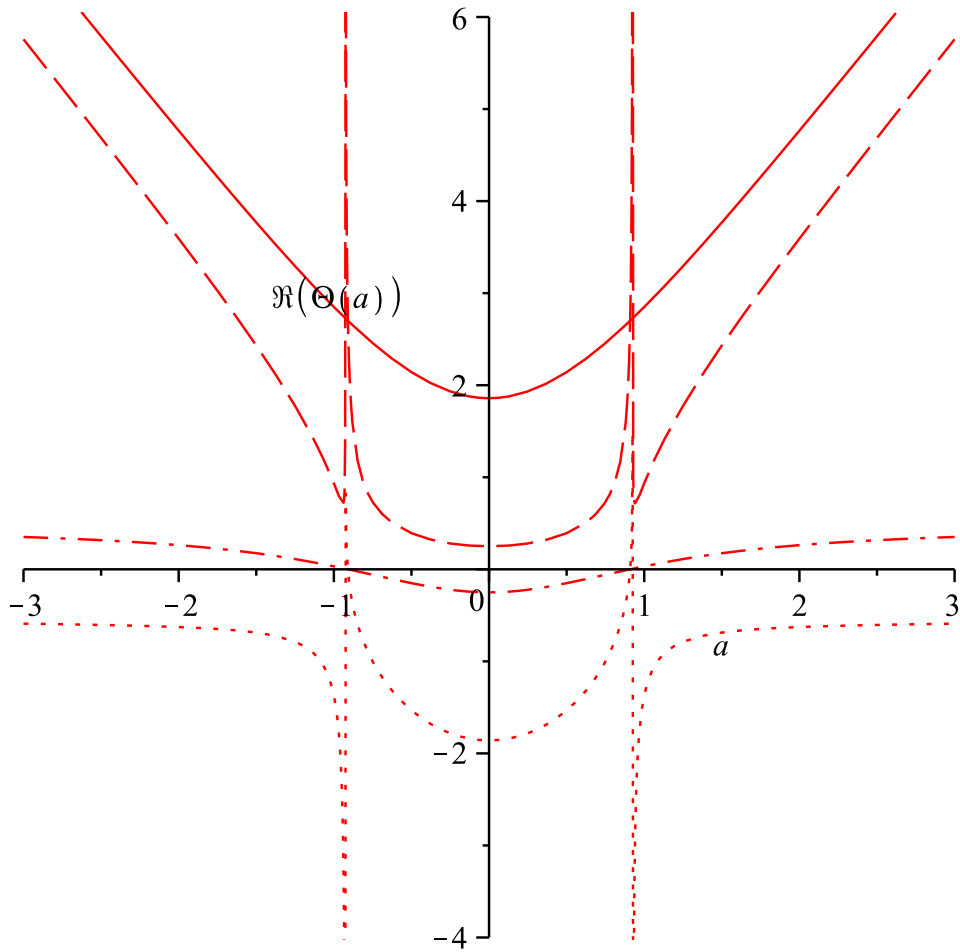}
\caption{\label{FIG_6_Papp}The $a$-dependence of real parts of precession angles for $B=6T$, $%
s=1$ (solid curve), $s=-1$ (dashed curves) and $K=i$. The $a$-dependence of $\Im\Theta (B=6,a;s=\pm 1,K=i$ ) has also been inserted, as shown by
dot-dashed ($s=1$) and dotted ($s=-1$) curves.
}
\end{figure}

\section{CONCLUSIONS}

In this paper the parameter dependence of novel spin-precession effects
proposed before\cite{9} has been discussed in some more detail. Such effects
concern equal coupling strength combinations of Rashba and Dresselhaus
spin-orbit interactions characterized by the dimensionless coupling $a$
under the influence of in-plane magnetic fields of magnitude $B$. The Zeeman
interaction has been favoured to the detriment of the orbital effects of the
magnetic field\cite{6}. A related discrete parameter, say $s=\pm 1$, has
also accounted for. The spin precession angle has been derived in terms of
suitable $\varepsilon $-expansions proceeding in terms of the convergence
condition $\varepsilon ^{2}<1$, which provides nontrivial manifestations.
This condition is fulfilled automatically when $s=1$, but we have to resort
to admissible regions in the parameter space such as given by (24) and (28)
in so far as $s=-1$, i.e. to $B\in I_{B}(a)$ and $a\in \widetilde{I}_{a}(B)$%
. To this aim parameter dependent crossing points between precession angles
for $s=1$ and $s=-1$, namely $B=$ $B_{\pm }(a)$ and $a=a_{\pm }(B)$, have
been established in an explicit manner. In addition, the precession angles
get characterized by nontrivial zeros like $B_{\pm }^{(0)}(a)$ and $a_{\pm
}^{(0)}(B)$, which are located specifically within admissible intervals. It
should be mentioned that the precession angles become imaginary within the
forbidden intervals $C_{B}(a)$ and $\widetilde{C}_{a}(B)$, which are
complementary to $I_{B}(a)$ and $\widetilde{I}_{a}(B)$, respectively.
Meaningless vanishingly small strips starting upwards from $\Theta =\Theta
_{1}$ and $\Theta =\Theta _{2}$, respectively, have been reasonably ignored.
Accordingly, the parameter dependence of the precession angles is
characterized by the interplay between admissible and forbidden intervals,
as illustrated by Figs. 1-4. Extrapolations of the wavenumber towards
imaginary and complex values have also been done, as illustrated by Figs. 5
and 6. Such structures may be useful for further developments. It is clear
that starting from admissible intervals, the parameters can be tuned until
the real part of the precession angle for $s=-1$ becomes zero:

\begin{equation}
\Theta (B,a;s=-1)\neq 0\rightarrow \Re\Theta (B,a;s=-1)=0
\end{equation}%
This proceeds via $B\in I_{B}(a)\rightarrow B\in $ $C_{B}(a)$ or \ $a\in
\widetilde{I}_{a}(B)\rightarrow a\in \widetilde{C}_{a}(B)$. Accordingly, the
$s=-1$-spin precession effects get ruled out. This means that such effects
are able to be switched off and on, which may serve to the description of
further manipulations. Actual zeros of the precession angles can also be
approached via $B\rightarrow B_{\pm }^{(0)}(a)$ and $a\rightarrow a_{\pm
}^{(0)}(B)$. In other words, one gets faced with nontrivial interplays
between admissible and forbidden regions, which stand for the main results
obtained in this paper. So we are in a position to emphasize that such
results are able to provide a deeper understanding of novel spin-precession
effects characterizing quantum wire models. Besides spin-filtering effects%
\cite{27} and transport properties\cite{28,29}, the
incorporation of dynamic localization effects\cite{30}, as well as of time
dependent magnetic fields\cite{31}, deserves further attention.

\appendix
\section{MATHEMATICAL DETAILS}

Resorting to orthonormalized spinors like

\begin{equation}
\left\vert \Psi _{\pm }\right\rangle =\left(
\begin{array}{c}
\psi _{1} \\
\pm \psi _{2}%
\end{array}%
\right)
\end{equation}%
let us introduce a spinorial representation for which $k\left\vert \Psi
_{\pm }\right\rangle =k_{\pm }\left\vert \Psi _{\pm }\right\rangle $. This
leads to

\begin{equation}
\exp (ikx)\left\vert \Psi _{\pm }\right\rangle =\exp (ik_{\pm }x)\left\vert
\Psi _{\pm }\right\rangle
\end{equation}%
where $k_{+}$ and $k_{-}$ stand for the selected wavenumber realizations
introduced above in accord with (25). The displaced wavefunction is then
given by

\begin{equation}
\left\vert \Psi (x)\right\rangle =G(x)\left\vert \Psi (0)\right\rangle
\end{equation}%
where

\begin{equation}
\left\vert \Psi \left( 0\right) \right\rangle =\left(
\begin{array}{c}
a \\
b%
\end{array}%
\right)
\end{equation}%
is an arbitrary normalized spinor, while

\begin{equation}
G(x)=\exp (ikx)\left( \left\vert \Psi _{+}\right\rangle \left\langle \Psi
_{+}\right\vert +\left\vert \Psi _{-}\right\rangle \left\langle \Psi
_{-}\right\vert \right) \quad .
\end{equation}%
Using (4) and (8), one finds the scalar product

\begin{equation}
\left\langle \Psi _{\pm }\mid \Psi (0)\right\rangle =\frac{1}{2}\left( a\pm
b\exp (i\beta \right)
\end{equation}%
so that a displacement of length $x=L$ along the $Ox$-axis is given by

\begin{equation}
\left\vert \Psi (L)\right\rangle =\exp (i\Omega _{0}L)M\left\vert \Psi
(0)\right\rangle
\end{equation}%
where $M$ denotes the unitary matrix

\begin{equation}
M=\left[
\begin{array}{cc}
\cos (\Omega _{1}L) & i\sin (\Omega _{1}L)\exp (i\beta ) \\
i\sin (\Omega _{1}L)\exp (-i\beta ) & \cos (\Omega _{1}L)%
\end{array}%
\right] \quad .
\end{equation}%
which proceeds in accord with Refs. 10 and 11. Accordingly, the precession
angle is given by

\begin{equation}
\Theta (B,a;s)=2\Omega _{1}L
\end{equation}%
which reproduces (25) via $\Omega _{1}=\varepsilon \Omega _{0}/2$.


\begin{thebibliography}{99}
\bibitem{1} R. Winkler, \textit{Spin Orbit Coupling Effects in
Two-Dimensional Electron and Hole Systems} (Springer, Berlin,
2003).

\bibitem{2} E. Rashba, Physica E \textbf{34}, 31 (2006).

\bibitem{3} G. Dresselhaus, Phys. Rev. \textbf{100}, 580 (1955).

\bibitem{4} E. Lipparini , M. Barranco, F. Malet, M. Pi and L. Serra,
Phys. Rev. B \textbf{74}, 115303 (2006).

\bibitem{5} R. G. Nazmitdinov, K. N. Pichugin and M. Valin-Rodriguez, Phys.
Rev. B \textbf{79}, 193303 (2009).

\bibitem{6} M. Scheid, I. Adagideli, J. Nitta and K. Richter, Semicond.
Science and Technology \textbf{24}, 064005 (2009).

\bibitem{7} M. Duckheim , D. Loss, M. Scheid, K.Richter, I.Adagideli and Ph.
Jacquod, Phys. Rev. B \textbf{81}, 085303 (2010).

\bibitem{8} J. Schliemann, J. C. Egues and D. Loss, Phys. Rev. Lett. \textbf{
90}, 146801 (2003).

\bibitem{9} C. Micu and E. Papp, Superlattices and Microstructures \textbf{
51}, 651 (2012).

\bibitem{10} V. M. Ramaglia, V. Cataudella, G. De Filippis and C. A. Perroni
, Phys. Rev. B \textbf{73}, 155328 (2006).

\bibitem{11} C. Cohen-Tannoudji, B. Diu and F. Lalo\"{e}, \textit{Quantum
Mechanics}, Vol. 2 (Wiley Interscience, New-York , 2006).

\bibitem{12} Ya. Zhang and F. Zhai, Phys. Rev. B \textbf{79}, 085311 (2009).

\bibitem{13} X. F. Wang and P. Vasilopoulos, Phys. Rev B \textbf{72
}, 085344 (2005).

\bibitem{14} M. Zarea and S. E. Ulloa, Phys. Rev. B \textbf{72}, 085342
(2005).

\bibitem{15} S. Q. Shen, Y. J. Bao, M. Ma, \ X. C. Xie and F. C. Zhang ,
Phys. Rev. B \textbf{71}, 155316 (2005).

\bibitem{16} M. Valin-Rodriguez and R. G. Nazmitdinov, Phys. Rev. B \textbf{73}
, 235306 (2006).

\bibitem{17} D. Zhang, J. Phys. A  \textbf{39}, L477 (2006).

\bibitem{18} N. N. Bogoliubov and Y.A. Mitropolsky, \textit{Asymptotic
Methods of Nonlinear Oscillations} ( Gordon and Breach, New-York, 1961).

\bibitem{19} S. Datta and B. Das, Appl. Phys. Lett. \textbf{56}, 665 (1990).

\bibitem{20} E. N. Bulgakov and A. F. Sadreev, Phys. Rev. B \textbf{66},
075331 (2002).

\bibitem{21} M. J. vanVeenhuizen, T. Koga and  J. Nitta, Phys. Rev. B \textbf{
73}, 235315 (2006).

\bibitem{22} A. Aharony, Y. Tokura, G. Z. Cohen, O. Entin-Wohlman and S.
Katsumoto, Phys. Rev. B \textbf{84}, 035323 (2011).

\bibitem{23} Q. P. Wu, X.D. He and Z. F. Liu,  Physica E  \textbf{44}, 738
(2011).

\bibitem{24} A. N. M. Zainuddin, S. Hong, L. Siddiqui, S. Srinivasan and S.
Datta, Phys. Rev. B \textbf{84}, 165306 (2011).

\bibitem{25} T. P. Pareek and P.Bruno, Phys. Rev. B \textbf{65}, 241305(2002).

\bibitem{26} O. E. Raichev and P. Debray, Phys. Rev. B \textbf{65}, 085319
(2002).

\bibitem{27} S. J. Gong and Z. Q. Yang, J. Appl. Phys. \textbf{102}, 033706
(2007).

\bibitem{28} J. E. Birkholz and V. Meden, J. Phys.:Condens. Matter \textbf{20
}, 085226 (2008).

\bibitem{29} L. Molenkamp and J. Nitta, Semicond. Sci. Technol.\textbf{24}
, 060301 (2009).

\bibitem{30} E. Papp and C. Micu, Physica E \textbf{44}, 1 (2011).

\bibitem{31} E. Papp, C. Micu and L. Aur, Superlattices and Microstructures
\textbf{44}, 770 (2008).
\end{thebibliography}
\end{document}